\documentclass[12pt]{article}
\usepackage{latexsym}
\usepackage{amsmath}
\usepackage{amssymb}
\usepackage{verbatim}
\usepackage{boxedminipage}
\newcommand{\xp}[1]{$^{\mbox{\scriptsize #1}}$}
\newcommand{\bibspace}{\vspace{1ex}\noindent}
\newcommand{\gap}{\noindent\hspace*{1cm}}%
\begin{document}
\begin{center}
\Large
\textbf{On the relation between quantum theory}\\
\textbf{and probability}\\[0.5cm]
\large
\textbf{Louis Marchildon}\\[0.5cm]
\normalsize
D\'{e}partement de chimie, biochimie et physique,
Universit\'{e} du Qu\'{e}bec,\\
Trois-Rivi\`{e}res (Qu\'{e}bec) G9A 5H7, Canada\\
\verb+louis.marchildon@uqtr.ca+
\end{center}
%
%
\begin{abstract}
The theory of probability and the quantum
theory, the one mathematical and the other physical,
are related in that each admits a number of very
different interpretations. It has been proposed
that the conceptual problems of the quantum theory
could be, if not resolved, at least mitigated by a
proper interpretation of probability. We
rather show, through a historical and analytical
overview of probability and quantum theory, that
if some interpretations of the one and the other
go along particularly well, none follows in a
unique way.
\end{abstract}
\section*{Outline}
\noindent
1. Introduction

\medskip\noindent
2. The interpretation of probability\\
\gap 2.1 The classical interpretation\\
\gap 2.2 The frequency interpretation\\
\gap 2.3 Propensity\\
\gap 2.4 The logical interpretation\\
\gap 2.5 The subjective interpretation

\medskip\noindent
3. Born's rule and measurement

\medskip\noindent
4. Indeterministic interpretations of quantum theory\\
\gap 4.1 The collapse of the state vector\\
\gap 4.2 The Copenhagen interpretation\\
\gap 4.3 The observer

\medskip\noindent
5. Deterministic interpretations of quantum theory\\
\gap 5.1 The theory of Bohm and de Broglie\\
\gap 5.2 Many worlds

\medskip\noindent
6. Conclusion
\section{Introduction}
The popularity of gambling dates back to ancient times,
as evidenced by bones and ivory artifacts found
by archaeologists and clearly used for this
purpose.\footnote{Hacking (1990, 2006) and Galavotti
(2005) recall key episodes in the history of probability.}
Yet although it gives the gambler an undeniable
advantage, the calculus of probability was not
developed until much later. In the Renaissance,
a \emph{probable} opinion was not understood as one
backed by evidence, but rather as one
that came from a recognized authority.

The concept of probability as we know it crystallized
in a few places in Europe around 1660, in particular
through the correspondence between Blaise Pascal and
Pierre de Fermat. Hacking (2006) notes that from the
beginning probability has been associated with two
distinct notions: one, subjective, consisting in a
degree of belief and the other, objective, referring
to random processes displaying in the long run
stable relative frequencies.

Beyond games of chance, probability theory quickly found
applications in the calculation of premiums for life insurance
and life annuities. In physics, however, and in spite of major
contributions by Pierre Simon de Laplace, probability theory
did not become truly important until the middle of the
nineteenth century, with the rise of statistical physics.

Developed in the second half of the nineteenth century
by James Clerk Maxwell, Ludwig Boltzmann and Josiah
Willard Gibbs, statistical physics aims to explain the
thermodynamic properties of a gas (for example) on
the basis of atomic and molecular theory. The theory
assumes that the gas is made up of a huge number of
molecules (typically, $10^{25}$) that move and
collide according to the deterministic laws of
Newtonian mechanics. Since, however, it is impossible
to solve the equations of motion of all these molecules,
one must resort to approximations. They consist in associating
to thermodynamic parameters (e.g.\ temperature, pressure,
energy and entropy) averages of microscopic parameters
computed under the assumption of a statistical model.
Several features of the model (for example, the
conservation of energy of an isolated system) can be justified
objectively. Yet the introduction of probabilities is in this
case only a second-best choice, linked to a lack of knowledge
of the exact conditions of the molecules. Paradoxically, this
compromise often allows to predict the values of certain
macroscopic parameters with remarkable accuracy.

Hacking (1990) points out that the concept of chance, associated
with a purely random process irreducible to deterministic laws,
really took off only at the end of the nineteenth century,
with Charles Sanders Peirce in particular. In physics, it was
first brought to light through the phenomenon of radioactivity.

Discovered in 1896 by Henri Becquerel, radioactivity was quickly
studied by Marie and Pierre Curie and by Ernest Rutherford.
The latter noted that what he called ``thorium emanation''
(our radon~220) displays an exponentially decreasing
radioactivity (Rutherford, 1900). Specifically, if there are
$N_0$ active atoms at time $t=0$, then the number of active
atoms at time $t>0$ is given by
\begin{equation}
N(t) = N_0 \exp \{ - \lambda t \}
\end{equation}
where $\lambda$ is a positive constant that characterizes
the radioactive substance. The exponential law implies that
the probability of a given atom emitting radiation does
not depend on time. How comes that the probability of radioactive
decay does not depend on the age of the atom? Why does one atom
decay now, and another much later? Here is how James Jeans
retrospectively summarized the situation:
\begin{quote}
In a milligram of radium, about 500 million
atoms disintegrate every second \mbox{[\ldots]}
Interesting but difficult questions arise
when we discuss which atoms will disintegrate
first, and which will survive longest without
disintegration \mbox{[\ldots]} [I]t seemed to
remove causality from a large part of our picture
of the physical world \mbox{[\ldots]} If we are
told the position and the speed of motion of
every [atom of radium] at any moment, we might
expect that Laplace's supermathematician would
be able to predict the future of every atom.
And so he would if their motion had conformed to
the classical mechanics. But the new laws merely
tell him that one of his atoms is destined to
disintegrate today, another tomorrow, and so on.
No amount of calculation will tell him which atoms
will do this. (Jeans, 1943, pp.~148--150)
\end{quote}

Although the question arose early on, most investigators did not
immediately conclude that the principle of causality had to be
abandoned. For Rutherford and Marie Curie, for example, complex
atomic processes, impossible to specify at the time, could be
at the origin of the apparent violation of causality,\footnote{See
Pais (1986), p.~123.} in the same way, perhaps, that the
randomness of the tossing of a die comes from not knowing
the exact conditions of the throw. This twofold way of
understanding radioactivity would be carried to the quantum
theory, which can be construed in a deterministic or
indeterministic way.

The purpose of this paper is to explore and clarify the link
between quantum theory and probability theory. We will first
recall, in section~2, different ways of interpreting
probability. Section~3 will be devoted to Born's rule, the key
to the introduction of probability in quantum theory,
and to the quantum measurement problem, the origin of the
theory's different interpretations. These interpretations can be
indeterministic or deterministic. We will examine, in
sections~4 and~5, how the interpretation of probability can be
adapted to them, before concluding in a more general way. 
%
\section{The interpretation of probability}
The distinction between objective and subjective
probability, already entertained by the founders of the theory,
as well as the gradual realization that natural phenomena
can in principle be necessary, contingent or purely random,
led to the development of different interpretations
of probability. Mellor (2005) classifies them in three types:
(i) physical probability, i.e.\ chance (e.g.\ the probability
that such and such radioactive nucleus will decay in the next
hour); (ii) epistemic probability (in statistical physics, in
particular, where it is related to the lack of knowledge of the
exact initial conditions); and (iii) an agent's belief with
respect to a contingent situation (which leads him, for example,
to bet 2 to 1 on the victory of a sports team). For her part,
Galavotti (2005) proposes another classification,
articulated historically under five types, which is well suited
to our analysis.\footnote{Galavotti (2005) provides many
references to the contributions of the various investigators
we will mention.}
\subsection{The classical interpretation}
From the beginning, probability theory has been concerned
with games of chance, such as the tossing of a die or the
throw of a coin. As long as the die is made properly,
none of the six faces is favored. We then say that the
probability of getting a specific outcome, say three,
is equal to 1/6, since three is one of the six possible
outcomes. This is the \emph{classical} conception of
probability, formulated by Laplace in his \emph{Philosophical
Essay on Probabilities}:
\begin{quote}
The theory of chance consists in reducing all the events
of the same kind to a certain number of cases equally
possible, that is to say, to such as we may be equally
undecided about in regard to their existence, and in
determining the number of cases favorable to the event
whose probability is sought. The ratio of this number
to that of all the cases possible is the measure of
this probability, which is thus simply a fraction
whose numerator is the number of favorable cases and
whose denominator is the number of all the cases possible.
(Laplace, 1814, pp.~6--7)
\end{quote}

Thus, Laplace considers that two cases are equally possible
if one is equally undecided about them. For him this
indecision, or lack of knowledge, is the only justification
for the use of probabilities. Indeed, Laplace considers that
``[a]ll events, even those which on account of their
insignificance do not seem to follow the great laws of nature,
are a result of it just as necessarily as the revolutions of
the sun'' (Laplace, 1814, p.~3). The universe obeys a rigorous
determinism. Probabilities are justified by our limited
knowledge of the laws of nature and of initial conditions,
and are therefore \emph{epistemic}.

The classical interpretation of probability has been the
object of several criticisms. It can be very difficult,
for example, to enumerate all the cases. \emph{A fortiori},
dividing them into ``equally possible'' cases already
implies, in a circular way, a prior notion of probability.
If, on the other hand, the number of cases is infinite,
the fraction that Laplace speaks of is indeterminate.
Subsequent interpretations will attempt to answer
these criticisms.
\subsection{The frequency interpretation}
Several names are associated with the frequency
interpretation of probability, in particular those
of Robert Leslie Ellis, John Venn, Richard von Mises,
Hans Reichenbach and Ernest Nagel. For these investigators,
the probability of a type of event refers to its relative
frequency in an arbitrarily long and, in the limit,
infinite sequence.

Not all sequences necessarily give rise to probabilities.
For von Mises, for example, the sequence must correspond
to a specific ensemble of events, such as the repeated
toss of a coin. It should also have a random character,
a concept of which we naturally have an intuition but
which is very difficult to define rigorously.

Among the objections made to the frequency interpretation,
two particularly stand out. The first consists in observing
that to identify a probability with a relative frequency,
one should strictly speaking consider an infinite sequence,
which is in practice impossible. The second is the difficulty
of associating, in terms of frequency, a probability to a
single event. One solution consists in considering the event
as belonging to a class. To evaluate, for example, the
probability that an individual will die in the coming
year, we consider the class of people of the same age,
same sex and comparable health. The specification of such
classes (``comparable health''), however, creates new
problems.

Many consider these objections serious and try to define
probability differently. Yet whatever the preferred way
of defining probability, testing a specific statistical
hypothesis can only be done by observing relative
frequencies.
\subsection{Propensity}
Anticipated by Charles Sanders Peirce, the propensity
interpretation of probability was really developed
by Karl Popper. He proposed it (i) to solve the problem
of the interpretation of quantum theory (Popper, 1959,
1982) and (ii) to define the probability of single
events (Popper, 1959).

Popper begins his analysis on the basis of the frequency
interpretation, which he had previously advocated.
From this point of view, asserting that the probability of
obtaining three, when throwing a die, is equal to 1/6 means
that the relative frequency of a three in a virtually infinite
sequence of throws is 1/6. Popper notes that the sequence is not
arbitrary, but corresponds to the specification of experimental
conditions that are repeated from one throw to the next.
The probability of getting three in a particular throw is thus
construed as being related to these experimental conditions.
Popper then proposes
\begin{quote}
\emph{a new physical hypothesis} (or perhaps a metaphysical
hypothesis) analogous to the hypothesis of Newtonian forces.
It is the hypothesis that every experimental arrangement
(and therefore every state of a system) generates physical
propensities which can be tested by frequencies.
(Popper, 1959, p.~38).
\end{quote}
Although not directly observable, propensities are nonetheless,
according to Popper, objective.

\begin{center}
\begin{boxedminipage}{12cm}
\begin{center}
\textbf{Bayes' Theorem}
\end{center}

\medskip\noindent
Let $A$ and $B$ be two random variables and let $a$
and $b$ be values of these variables. The
\emph{conditional probability} $P(a|b)$, i.e.\ the
probability that $A$ has value $a$ if $B$ has value
$b$, is defined as $P(a,b)/P(b)$, where $P(a,b)$ is
the probability that $A$ has value $a$ and $B$ has
value $b$. Because $P(a,b) = P(b,a)$, we immediately
obtain
\[
P(b|a) = P(a|b) P(b) / P(a)
\]
This is \emph{Bayes' theorem}. It is especially useful
to evaluate the probability of a hypothesis conditional
to an observation.

\medskip
As an example, suppose that $A$ is associated with the
outcome of the throw and $B$ with the choice of die
(see below). Let $+$ represent the fair die and $-$ the
rigged die. Then
\[
P(-|3) = P(3|-) P(-) / P(3)
\]
By assumption, $P(-) = 1/2 = P(+)$. This means that
$P(3) = 5/24$, the average of 1/4 and 1/6. Since
$P(3|-) = 1/4$, we find that $P(-|3) = 3/5$.
\end{boxedminipage}
\end{center}

It is important to note that, although propensity can be used
to shed light on the notion of probability, the two concepts
cannot be identified. Consider two dice, the first fair
and the second rigged so that the probability of obtaining three
by throwing it is 1/4. It is natural to quantify the propensity
of getting three by throwing each die. On the other hand,
Bayes' theorem tells us that if the \emph{a priori} probability
of throwing each die is equal, and we get three, then the
probability of having thrown the rigged die is equal to 3/5
(see box). In this case, it would be highly artificial to speak
of the propensity to use the rigged die. To associate probability
with propensity, there must be a causal link between the
experimental conditions and the outcome.
\subsection{The logical interpretation}
Anticipated by Auguste de Morgan and George Boole, this point
of view was developed in the twentieth century, especially
by John Maynard Keynes and Rudolf Carnap.

The basic idea of the logical interpretation is to associate
probability with a degree of belief. Unlike the subjective
interpretation that we shall introduce in the next section,
we do not have in mind here the belief of specific agents, but
rather the belief of an ideal rational agent. Probability
is thus a logical relation between propositions, referring
to arguments that do not lead to a unique conclusion.

The abstract nature of the logical interpretation makes it
difficult to apply it directly to real situations. This led
Carnap to propose two distinct notions of probability.
Probability$_1$, a logical concept, represents the degree of
confirmation brought to a given hypothesis by different clues.
Probability$_2$, on the other hand, refers to relative frequencies.
The two concepts are related inasmuch as the first kind of
probability can be used to estimate the second one.
\subsection{The subjective interpretation}
The identification of probability with a degree of belief
characterizes the subjective interpretation, developed
in the last century mainly by Frank Ramsey and Bruno de
Finetti. That belief, according to these investigators,
is in no way unique, nor is it determined by considerations
of rationality. Two distinct agents can have distinct degrees
of belief, with the only restriction that the beliefs of each
must be consistent.

De Finetti proposes an operational definition of probability in
terms of bets. Believing that an event will occur once in every
four times, for example, means that the agent is willing to bet
one dollar against three on its occurrence. Consistency
consists in restricting the type of bets so that no combination
leads to a sure loss.

Of course, observation and experience imply that an agent's
beliefs can change. For example, suppose I believe that a
die is fair, and therefore that each of the six possible
outcomes of a throw has the same probability. If I observe that
in 100 subsequent throws, the outcome three occurs 25 times,
my initial belief will be altered. Bayes' rule allows me to
refresh my belief. Thus two agents with different initial
beliefs may end up with closer beliefs. Nevertheless this
does not mean, according to de Finetti, that the beliefs
converge to a probability that would be objective.
%
\section{Born's rule and measurement}
The formalism of quantum theory was developed in 1925--1926
by Werner Heisenberg, Erwin Schrödinger and Paul Dirac.
As early as 1926, Max Born brought to light the probabilistic
nature of the theory.

The formalism of the theory can be interpreted in several ways.
We will see that each interpretation of the theory naturally
adapts to an interpretation of probability.

In this section, we will point out elements of the
quantum theory that do not depend on the theory's
interpretation. Such is, in particular,
Born's rule, a statistical postulate that introduces
probability in a purely operational way.

Any physical system adequately described by quantum theory
will be called a \emph{quantum system}. Many believe that
quantum theory has universal scope, and therefore
that all physical systems are quantum systems. It is not
necessary at this time to commit oneself on this issue.
Everyone agrees, however, that atoms and simple molecules
are quantum systems. In what follows, the term \emph{atomic
system} will denote any microscopic physical system that
obeys the laws of quantum theory.

The description of a quantum system is carried out by means
of a mathematical object called a \emph{state vector}.
The most general description is done by means of a
\emph{density operator} (or \emph{density matrix}), which
doesn't need to be introduced at this stage. The Schr\"{o}dinger
\emph{wave function} is an example of state vector. It is
typically denoted by means of a \emph{ket} like $|\phi\rangle$. The
interpretation of the state vector is a controversial issue.
Some believe that it describes an individual system, others a
statistical ensemble of systems, while for others still it
represents an agent's information about the system. Whatever the
preferred interpretation, however, the state vector corresponds
to a \emph{preparation} procedure. An example of preparation
consists in aiming photons of a given wavelength toward
a linear polarizer. All photons from the polarizer are then
associated with a specific state vector. The preparation process
constitutes an operational definition of the state vector,
which somehow isolates its uncontroversial features.

The state vector is an element of a vector space $\mathcal{V}$
that is called the \emph{state space}. Although the dimension
of the state space of many quantum systems is infinite, we can
restrict ourselves to finite-dimensional spaces. In this case,
$\mathcal{V}$ coincides with $\mathcal{C}^N$, the complex vector
space of dimension~$N$, in which the usual scalar product is
defined. All identical quantum systems (for example, all helium
atoms) have isomorphic state spaces. And each vector of the state
space corresponds, in principle, to a possible preparation
procedure. The correspondence, however, is not one-to-one.
Two vectors which are multiples of each other correspond to the
same preparation, i.e.\ to the same physical situation (to the
same state of the system, if the vector is so
interpreted).\footnote{It may also happen that two
operationally distinct procedures prepare the same quantum
state.}

Let $\{|a_i\rangle, i = 1\ldots N\}$ be an orthonormal basis of
the state space. It is always possible, in principle, to construct
a macroscopic measuring apparatus that has the following property.
If the quantum system is prepared in the state $|a_i\rangle$,
then at the end of the measurement the apparatus indicates
the value $\alpha_i$, where these values are all distinct real
numbers. Many interpret this situation by stating that the
apparatus then measures a physical quantity $A$ associated with
the atomic system, whose possible values are the $\alpha_i$.

The vectors $|a_i\rangle$ make up a basis of the state space.
Therefore, any normalized vector $|\phi\rangle$ can be
expressed as a linear combination of the $|a_i\rangle$,
that is,
\begin{equation}
|\phi\rangle = \sum_{i=1}^N c_i |a_i\rangle
\label{superp}
\end{equation}
where the $c_i$ are complex numbers. Operationally, \emph{Born's
rule} is then formulated as follows. If one prepares a large number
of identical systems in state $|\phi\rangle$, and performs on each
system the measurement specified above, then one will obtain the
value $\alpha_j$ with relative frequency $|c_j|^2$ (in the
limit where the number of systems tends to infinity). Formulated
in this way, the rule depends neither on the interpretation of
quantum theory nor on that of probability.

Born's rule refers to the notions of measurement and apparatus.
From a phenomenological point of view, these notions seem fairly
clear. If, however, we wish to describe the measurement
process in more detail, we have to be more precise.

In the remainder of this section, we will assume that the state
vector of a quantum system describes an individual system
(rather than, for example, a statistical ensemble of systems).
From a fundamental point of view, an apparatus consists of
a very large number of atoms, arranged in a complex way.
Individually as well as in restricted aggregates, these atoms
obey the laws of quantum theory. Let us assume that quantum
theory is truly fundamental and that the scope of its
applications has no limit. In this case, the measuring apparatus
itself constitutes a quantum system, to which we can associate a
state space. It is, of course, a very complicated space,
but the sole hypothesis of its existence leads to unexpected
conclusions.

To perform a measurement of the physical quantity $A$, the
apparatus must be able to display $N$ different values $\alpha_i$.
To do this, the state space of the apparatus must contain at
least $N$ orthogonal vectors $|\alpha_i\rangle$, each vector
corresponding to a state where the apparatus displays the
corresponding value.\footnote{As a matter of fact, there is a
very large number of state vectors associated with each value
$\alpha_i$, but this will not change our conclusion.}
We can also assume that there is a vector $|\alpha_0\rangle$
that corresponds to an initial state where the apparatus does
not indicate any value.

Consider a situation where the apparatus is prepared in state
$|\alpha_0\rangle$ and the atomic system is prepared in state
$|a_i\rangle$. We then say that the global system (consisting
of the atomic system and the apparatus) is prepared in state
$|a_i\rangle |\alpha_0\rangle$.\footnote{The object
$|a_i\rangle |\alpha_0\rangle$ is what is called a
\emph{tensor product}. This notion,
and others succinctly introduced here, are specified
in quantum mechanics textbooks, e.g.\ Marchildon (2000).}
If the atomic system is aimed toward the apparatus, the latter,
at the completion of the measurement, will indicate the
value $\alpha_i$. This implies (assuming the state vector of the
atomic system does not change) that the state vector of the global
system will be given by $|a_i\rangle |\alpha_i\rangle$.

What will happen now if we prepare the atomic system in the state
$|\phi\rangle$, represented by equation~(\ref{superp})?
The global system's initial state will then be given by
$|\phi\rangle |\alpha_0\rangle$. What will be the global system's
state at the end of the measurement? The dynamics of a quantum
system can be complicated but, for all microscopic systems it is
governed by a linear equation (the Schr\"{o}dinger equation,
in the nonrelativistic case). Assuming that this property is
universal, the evolution equation of the global system should
also be linear. In other words, the initial state vector
\begin{equation}
|\phi\rangle |\alpha_0\rangle
= \sum_{i=1}^N c_i |a_i\rangle |\alpha_0\rangle
\label{initial}
\end{equation}
should evolve in such a way that at the end of
measurement it becomes
\begin{equation}
\sum_{i=1}^N c_i |a_i\rangle |\alpha_i\rangle
\label{final}
\end{equation}

At first glance, this result seems quite different from what was
expected, i.e.\ a situation where the apparatus displays a
well-defined outcome, e.g.\ $\alpha_j$, with relative frequency
$|c_j|^2$. Nevertheless, this result is an inescapable consequence
of associating the state vector with an individual system and of
assuming that quantum theory has universal scope.
The incompatibility, apparent at least, of result~(\ref{final})
with what we would like to obtain is called the \emph{measurement
problem}. The various interpretations of quantum theory aim
primarily at solving this problem.
%
\section{Indeterministic interpretations of quantum theory}
From its beginning, and during the two or three decades that
followed, quantum theory was almost unanimously interpreted
in an indeterministic way. The Copenhagen interpretation, a
constellation of ideas proposed mostly by Bohr and also by
Heisenberg, ruled virtually
unchallenged for a long time (Freire Jr.,
2015). The expositions of quantum theory grafted onto it the
notion of state vector collapse, now recognized as largely
alien to Bohr's ideas. From the Copenhagen interpretation the
idea more recently developed that the state vector does not
represent the state of an atomic system, but the
information of a more or less ideal observer.
\subsection{The collapse of the state vector}
Considered very early by Dirac, the collapse of the state
vector was formalized by John von Neumann (1932). The idea is
to assume that the evolution of an atomic system does not
always follow a linear equation.

Specifically, von Neumann assumed that an atomic system can
evolve in two ways, which he called \emph{Process~1}
and \emph{Process~2}. Process~2 applies to every circumstance
other than a measurement, and it simply consists in the linear
evolution of the state vector, governed by the Schr\"{o}dinger
equation. Process~1, on the other hand, applies exclusively to
a measurement situation, and occurs as a result of the
transformation of vector~(\ref{initial}) into
vector~(\ref{final}). It consists in the transformation of
vector~(\ref{final}) into one of its terms, in a purely random
way. Von Neumann postulates, however, that the probability that
vector~(\ref{final}) transforms into the term
$|a_j\rangle |\alpha_j\rangle$ is equal to $|c_j|^2$.

It is easy to see that von Neumann's hypothesis implies Born’s
rule. In both formulations, we take for granted that the
notions of measurement and apparatus are clear enough.

In the collapse theory, the vector $|a_j\rangle |\alpha_j\rangle$
into which the superposition~(\ref{final}) transforms is
completely random. The statistical distribution of the outcomes,
however, depends entirely (through the
coefficients $c_i$ and the normalized vectors $|a_i\rangle$)
on the preparation of the atomic system and the nature of the
apparatus. In other words, it leads to objective probabilities
that depend entirely on the experimental setup. This is
precisely the context for which Popper proposed the
propensity interpretation.

Originally, von Neumann did not suggest any specific mechanism
through which collapse could occur. Such mechanisms came later,
one of the best known being \emph{spontaneous localization}
(Ghirardi, Rimini, \& Weber, 1986; Ghirardi, Pearle, \&
Rimini, 1990). This is a random process which, from time to time,
reduces the spatial extension of the wave function of a
particle. The theory is designed in such a way that the
localization of an atomic object is extremely rare, whereas
a macroscopic object (such as the pointer of an apparatus)
is localized in a few nanoseconds. As in von Neumann's approach,
probability is objective and neatly fits in the
propensity interpretation.

The collapse of the state vector aims at solving the measurement
problem. This has been formulated under the assumption that the
state vector describes an individual quantum system. What happens
if we do away with this assumption? Suppose, for example, that the
state vector describes not an individual system, but a statistical
ensemble of similarly prepared systems (Ballentine, 1970).
It is tempting then to assume that, at the end of the measurement,
the vector~(\ref{final}) represents not a superposition of
macroscopically distinct states, but a statistical ensemble
of systems in each of which the apparatus displays a specific
value. And could it not be that this value corresponds to the one
that the physical quantity would have had just before the
measurement, so that the latter would consist in a process
of separation of the systems of the statistical ensemble according
to the initial value of the physical quantity? Unfortunately,
this last hypothesis is untenable. It can indeed be shown
(Kochen \& Specker, 1967) that it is impossible to assign precise
values (even unknown ones) to all the quantities of a physical
system, if they satisfy the algebraic relations prescribed by
quantum theory. Nevertheless, it is possible to assign precise
values to some quantities, as we will see in section~5.1 with
hidden variables.
\subsection{The Copenhagen interpretation}
Bohr's and Heisenberg's ideas about the interpretation of
quantum theory were developed in the 30 years following the
advent of the theory.

According to Bohr and Heisenberg, an atomic system has
well-defined properties only in the context of a measurement,
carried out by means of an apparatus necessarily described
by the classical theory. Thus the problem of measurement stated
above does not arise, since one cannot attribute a state vector
to the apparatus. The problem of measurement gives way to
the problem of the distinction between the classical and the
quantum: for example, under what circumstances, or from what
size, mass or level of complexity does an aggregate of atoms
or molecules satisfy the laws of the classical theory rather
than those of the quantum theory?

The Heisenberg uncertainty principle, a pillar of the
Copenhagen interpretation, asserts that no state vector allows
to predict both the result of the measurement of the position
and the result of the measurement of the momentum of a particle
such as an electron. Moreover, no apparatus can simultaneously
measure both quantities. According to Bohr, this in no way
implies that the quantum theory is incomplete:
\begin{quote}
Although the phenomena in quantum physics can no longer be
combined in the customary manner, they can be said to be
complementary in that sense that only together do they exhaust
the evidence regarding the objects, which is unambiguously
definable. (Bohr, 1998, p.~130).
\end{quote}
Thanks to complementarity, the quantum theory gives a complete
and objective description of an atomic system.

The objective character of the description was also emphasized
by Heisenberg:
\begin{quote}
The probability function \mbox{[\ldots]} contains statements
about possibilities or better tendencies (``potentia''
in Aristotelian philosophy), and these statements are
completely objective, they do not depend on any observer.
(Heisenberg, 1958, p.~27).
\end{quote}

As Popper himself noted, Heisenberg's ``tendencies''
are akin to propensity. The probabilities depend only on the state
vector and the experimental configuration. Thus, the Copenhagen
interpretation fits well within the propensity interpretation of
probability.\footnote{It should be pointed out here that
Heisenberg's text quoted above goes on as follows:
``[The probability function also contains] statements about our
knowledge of the system, which of course are subjective in so far
as they may be different for different observers. In ideal cases
the subjective element in the probability function may be
practically negligible as compared with the objective one.
The physicists then speak of a `pure case'.''
This remark can apply to situations where the observer might
have only a partial knowledge of the preparation of the state
of the atomic system, which he must then describe by a density
operator. Heisenberg's remark can also anticipate the discussion
in the next section.}
\subsection{The observer}
In our presentation, state vector collapse occurs when a
physical quantity associated with an atomic system is
measured with a macroscopic apparatus. Several investigators
(London \& Bauer, 1939; Wigner, 1961) have suggested that the
collapse occurs at the moment a conscious subject becomes aware
of the outcome. This idea is less popular today, but the
intuition that an observer is necessary, if not for physical
collapse, at least for the formulation of the theory, is alive
and well. It was formulated in a precise way by Rudolph Peierls:
\begin{quote}
In my view the most fundamental statement of quantum mechanics
is that the wavefunction, or more generally the density matrix,
represents our \emph{knowledge} of the system we are trying
to describe. (Peierls, 1991, p.~19).
\end{quote}
Peierls believes that this assumption allows to view collapse
in a completely different light. Collapse is no longer a physical
process, but corresponds rather to a change in information.
There is therefore no reason for it to obey the Schr\"{o}dinger
equation.

What is the nature of the observer we have in mind here?
Of course, different concrete individuals can have more or less
complete knowledge about the atomic system. Is there an ideal
observer, whose knowledge would be maximal and more correct
than that of all the others? Although Peierls does not answer
this question precisely, his argument seems to call more for a
positive answer. If so, the logical interpretation of probability
seems to best represent Peierls' view.

Others, however, do not accept the idea of an ideal observer.
This is the case of \emph{QBism}, or \emph{quantum Bayesianism}.
According to this point of view,
\begin{quote}
quantum mechanics is a tool anyone can use to evaluate, on the
basis of one's past experience, one's probabilistic expectations
for one's subsequent experience. (Fuchs, Mermin, \& Schack,
2014, p.~749).
\end{quote}

Unlike in the Copenhagen interpretation, there is no distinction
here between the classical and the quantum. The distinction lies
between the agent and the rest of the world. Everything outside
an agent~A (including other agents) constitutes a quantum system
for agent~A. It is the same for any other agent. Different agents
generally have different beliefs about a given atomic system.
Thus
\begin{quote}
probabilities are assigned to an event by an agent \mbox{[\ldots]}
and are particular to that agent. The agent's probability
assignments express her own personal degrees of belief about
the event. (Fuchs, Mermin, \& Schack, 2014, p.~749).
\end{quote}
Therefore,
\begin{quote}
the QBist position, that quantum states are personal judgments
of an agent, is an inevitable consequence of the subjective
view of probability expressed so eloquently by Bruno de Finetti.
(Fuchs, Mermin, \& Schack, 2014, p.~753).
\end{quote}

Note in closing the pragmatic, or instrumentalist, character
of Peierls' and QBism's views. Their critics believe that these
approaches do not provide a sufficiently realistic description
of quantum systems and therefore fail to solve the measurement
problem.
%
\section{Deterministic interpretations of quantum theory}
\subsection{The theory of Bohm and de Broglie}
The formalism of quantum theory, based on the Schr\"{o}dinger
equation and the Born rule, does not allow to predict with
certainty the result of the measurement of a physical quantity.
Yet Bohr and Heisenberg have consistently argued that the
formalism is complete. If one cannot predict the result of the
measurement of a quantity it is, they claimed, because the
quantity does not have a well-defined value before the
measurement.

Quite early, however, other investigators (Louis de Broglie
and Albert Einstein in particular) adopted a different
attitude. According to them, the reason why quantum theory
does not make unique predictions is that it gives only an
incomplete description of an atomic system. In principle, so
they say, a more complete description of the system is possible.
The complete description would include, in addition to the
state vector, various parameters whose knowledge would allow
to make unique predictions. These are usually called
\emph{hidden variables}, since their values do not follow
from those of the state vector.

As early as 1927, de Broglie proposed a theory of hidden
variables. It really took off in 1952, when David Bohm
answered the objections to which it had given rise. This theory
postulates the existence of a preferred physical quantity,
in this case position. An electron, for example, always has a
well-defined position which, however, cannot be known
accurately. Position is the theory's hidden variable and,
together with the state vector, it gives a complete description
of the quantum system.

Let us see briefly how, for a particle like the electron, the
theory of Bohm and de Broglie is formulated.\footnote{Strictly
speaking, the theory of a single particle must be derived
from the theory of all particles in the universe, as explained
in detail by D\"{u}rr, Goldstein, \& Zanghí (1992).} The dynamics
of the electron is governed by two equations: on the one hand
the Schr\"{o}dinger equation, which determines the temporal
evolution of the wave function $\psi(\vec{r}, t)$; on the other
hand the equation of motion, which is given by
\begin{equation}
m \vec{v} = \hbar \vec{\nabla} S
\label{traj}
\end{equation}
Here $m$ is the mass and $\vec{v}$ is the velocity of the
electron, while $S$ is the phase of the complex wave function.
The trajectory of the electron is entirely determined by the
wave function, which de Broglie called the \emph{pilot wave}.

How can we, by means of a perfectly deterministic theory,
recover the statistical predictions of the quantum theory?
The answer is related to the state preparation process.
The Bohm and de Broglie theory assumes that the preparation
of an electron in the $\psi$ state is incompatible with the
specification of its position. Specifically, the theory assumes
that of the position of an electron in the $\psi$ state, we
only know the probability density, equal to $|\psi(\vec{r},t)|^2$.
Clearly, we are dealing here with a purely subjective probability:
the theory is deterministic, but our lack of knowledge of the
initial conditions compels us to use probabilities. These
probabilities, however, do not depend on the subjectivity
of a particular agent. They are related to an ideal agent
who knows the wave function exactly. They are thus associated
with the logical interpretation of probability.

It can be shown that the hypothesis that the position is
distributed according to the absolute square of the wave
function exactly reproduces all the statistical predictions
of the quantum theory. The hypothesis is also true at any time
if it is true at a given time. Specifically, if the position is
distributed according to $|\psi(\vec{r}, t_0)|^2$ at time
$t_0$, then equation~(\ref{traj}) and the Schr\"{o}dinger
equation imply that it will be distributed according to
$|\psi(\vec{r}, t)|^2$ at time~$t$.

Note finally that Bohm's and de Broglie's theory solves the
measurement problem without appealing to an assumption
such as collapse. At the end of the interaction between the
atomic system and the apparatus, the state vector is well
represented by equation~(\ref{final}). Nevertheless, the
position of the atomic system and those of the particles of
the apparatus are always well-defined, and are concentrated
in only one of the terms of~(\ref{final}). For all practical
purposes, one can account for the subsequent evolution of
the global system by retaining only this one term.
\subsection{Many worlds}
In the theory of Bohm and de Broglie, the wave function of a
quantum system always evolves unitarily. At the end of a
measurement, however, all but one of the terms of the
superposition vanish where the particles are located.

In a very different approach, Hugh Everett also proposed,
in 1957, that every quantum system always evolves unitarily.
Thus, there is no physical process that, just like collapse,
would transform a system described by~(\ref{final}) into
a system described by only one term of the sum.
Rather, Everett boldly assumes that all the terms of the
sum correspond to real systems.

Let's see more precisely what this means. An apparatus
measures a physical quantity associated with an atomic system,
a quantity that can take $N$ distinct values. Just before the
measurement, there is a system, an apparatus and, say,
a human observer, all of which, according to Everett, are
described by quantum theory. At the end of the measurement
there will be $N$ systems, $N$ apparatus (each pointing
to a distinct outcome) and $N$ observers (each observing
a distinct outcome). In other words, the initial world has
split into $N$ different worlds, any one as real as any
other.

What we have just described is called the \emph{many-worlds}
theory. Not everyone interprets Everett's hypothesis in such a
radical way. Some consider rather a splitting of the observer's
consciousness, others the formation of decoherent sectors of
the wave function. In fact, the nature of multiplicity is a
significant problem in Everett's approach (Marchildon, 2015).
We will focus on the many-worlds theory because it is
particularly clear, and lends itself well to probability
analysis.

At first glance, probability seems completely foreign to Everett's
approach. Everything, in fact, seems certain. Before the measurement,
the observer notices that the apparatus is in the neutral position.
If she knows quantum theory, and believes in Everett's approach,
she is certain that she will split into $N$ copies of herself,
each one recording that the apparatus indicates a specific value.
After measurement, each copy of the observer (say $O_j$) will note
that the apparatus indicates the corresponding value
(here $\alpha_j$).

Nevertheless, probability can be introduced by the following
argument, proposed by Lev Vaidman (1998). Suppose that throughout
the measurement process the observer is in a deep slumber.
She wakes up only when the measurement is completed, and does
not immediately take note of the value indicated by the
apparatus. At this moment, she does not know in which of the
many worlds she is. She doesn't know what is the value indicated
by the apparatus located in the same world as her. To the
question ``What value does your apparatus indicate?'' she cannot
give a categorical answer. She can only say (if she knows
quantum theory) that the probability that the apparatus indicates
the value $\alpha_j$ is equal to $|c_j|^2$. When she later
notes the result, she will of course have to correct her
judgment.

Thus, probability also finds a place in Everett's approach.
Just as in the theory of Bohm and de Broglie, we are dealing
with a subjective probability. As it is the judgment of an ideal
observer, the logical interpretation of probability is also
appropriate here.
%
\section{Conclusion}
There is no doubt that, among all physical theories, quantum
theory is the one where problems of interpretation are the
most acute. Probability theory is also characterized, in the
field of mathematics, by the diversity of its interpretations.
The problem, however, arises in different ways in the two
cases.

The main purpose of an interpretation of quantum theory is to
clarify how the formalism can account for the measurement of
properties of an atomic system. These interpretations all make
distinct assumptions and in that sense are mutually contradictory.
If the state vector collapses in the von Neumann manner, it cannot
always evolve in a unitary way as in the case of the Bohm and de
Broglie theory. And if all the results of a measurement coexist,
there can be no univocal quantity like position. The interpretations
can all try to answer the question ``How can the world be for the
quantum theory to be true?", but these answers are mutually
exclusive, and only one can correspond to the real world.
By contrast, there is no need to give a single interpretation of
probability theory.
\begin{table}[hbt]
\begin{center}
\begin{tabular}{||l|l||}\hline\hline
\textbf{Quantum theory}
& \textbf{Probability}\\ \hline 
Collapse & Propensity\\
Copenhagen & Propensity\\
Epistemic (Peierls) & Logical\\
QBism & Subjective\\
Bohm and de Broglie & Logical\\
Many worlds & Logical\\
\hline\hline
\end{tabular}
\caption{The relation between the interpretation of
quantum theory and the interpretation of probability}
\label{tab1}
\end{center}
\end{table}

In this article, we have examined different interpretations of
quantum theory and have pointed out, in each case, the interpretation
of probability that seems most adequate. These relationships are
summarized in Table~1. We will not claim that the interpretation
of probability solves the conceptual problems of quantum theory.
However, the clarification of the links between one and the other
allows to state them in a clearer way.
\section*{Acknowledgements}
I thank Frank Crispino, Marian Kupczynski and Jean-Ren\'{e}
Roy for their comments on an earlier version of the manuscript.
%
\section*{References}
Ballentine, L. E. (1970).
The Statistical Interpretation of Quantum Mechanics.
\textit{Reviews of Modern Physics},
42(4), 358--381.\\
\noindent\small
\verb+https://journals.aps.org/rmp/abstract/10.1103/RevModPhys.42.358+
\normalsize

\bibspace
Bohm, D. (1952).
A Suggested Interpretation of the Quantum
Theory in Terms of `Hidden' Variables, I and II.
\textit{Physical Review}, 85(2), 166--193.\\
\noindent\small
\verb+https://journals.aps.org/pr/abstract/10.1103/PhysRev.85.166+\\
\noindent
\verb+https://journals.aps.org/pr/abstract/10.1103/PhysRev.85.180+
\normalsize

\bibspace
Bohr, N. (1998).
\textit{Causality and Complementarity. The Philosophical
Writings of Niels Bohr, Vol.~IV} (edited by
J. Faye \& H. J. Folse).
Ox Bow Press.

\bibspace
De Broglie, L. (1927).
La m\'{e}canique ondulatoire et la structure
atomique de la mati\`{e}re et du rayonnement.
\textit{Le journal de physique et le radium},
8(5), 225--241.\\
\noindent\small
\verb+https://doi.org/10.1051/jphysrad:0192700805022500+
\normalsize

\bibspace
D\"{u}rr, D., Goldstein, S., \& Zangh\'{i}, N. (1992).
Quantum Equilibrium and the Origin of
Absolute Uncertainty.
\textit{Journal of Statistical Physics},
67(5/6), 843--907.\\
\noindent\small
\verb+https://link.springer.com/article/10.1007/BF01049004+
\normalsize

\bibspace
Everett, H. (1957).
`Relative State' Formulation of
Quantum Mechanics.
\textit{Reviews of Modern Physics},
29(3), 454--462.\\
\noindent\small
\verb+https://journals.aps.org/rmp/abstract/10.1103/RevModPhys.29.454+
\normalsize

\bibspace
Freire Junior, O. (2015).
\textit{The Quantum Dissidents: Rebuilding the Foundations
of Quantum Mechanics (1950--1990)}.
Springer.

\bibspace
Fuchs, C. A., Mermin, N. D., \& Schack, R. (2014).
An Introduction to QBism with an Application
to the Locality of Quantum Mechanics.
\textit{American Journal of Physics},
82(8), 749--754.\\
\noindent\small
\verb+https://doi.org/10.1119/1.4874855+
\normalsize

\bibspace
Galavotti, M. C. (2005).
\textit{Philosophical Introduction to Probability}.
CSLI Publications.

\bibspace
Ghirardi, G. C., Rimini, A., \& Weber, T. (1986).
Unified Dynamics for Microscopic and
Macroscopic Systems.
\textit{Physical Review D},
34(2), 470--491.\\
\noindent\small
\verb+https://journals.aps.org/prd/abstract/10.1103/PhysRevD.34.470+
\normalsize

\bibspace
Ghirardi, G. C., Pearle, P., \& Rimini, A. (1990).
Markov Processes in Hilbert Space and
Continuous Spontaneous Localization of Systems
of Identical Particles.
\textit{Physical Review A},
42(1), 78--89.\\
\noindent\small
\verb+https://journals.aps.org/pra/abstract/10.1103/PhysRevA.42.78+
\normalsize

\bibspace
Hacking, I. (1990).
\textit{The Taming of Chance}.
Cambridge University Press.

\bibspace
Hacking, I. (2006).
\textit{The Emergence of Probability} (2\xp{nd} ed.).
Cambridge University Press.

\bibspace
Heisenberg, W. (1958).
\textit{Physics and Philosophy. The Revolution
in Modern Science}.
Harper Perennial, 2007.

\bibspace
Jeans, J. (1943).
\textit{Physics and Philosophy}.
Cambridge University Press.

\bibspace
Kochen, S., \& Specker, E. P. (1967).
The Problem of Hidden Variables in
Quantum Mechanics.
\textit{Journal of Mathematics and Mechanics},
17(1), 59--87.\\
\noindent\small
\verb+https://link.springer.com/chapter/10.1007/978-94-010-1795-4_17+
\normalsize

\bibspace
Laplace, P. S. (1814).
\textit{A Philosophical Essay on Probabilities}.
(English translation from the 6\xp{th} ed.\
by F. W. Truscott \& F. L. Emory.).
Wiley, 1902.\\
\noindent\small
\verb+https://archive.org/details/philosophicaless00lapliala/page/n5/mode+\\
\verb+/2up+
\normalsize

\bibspace
London, F., \& Bauer, E. (1939).
La th\'{e}orie de l'observation en mécanique quantique.
\textit{Actualit\'{e}s scientifiques et industrielles}, 775.
Hermann.

\bibspace
Marchildon, L. (2000).
\textit{Quantum Mechanics: From Basic Principles to
Numerical Methods and Applications}.
Springer, 2002.

\bibspace
Marchildon, L. (2015).
Multiplicity in Everett's Interpretation
of Quantum Mechanics.
\textit{Studies in History and Philosophy
of Modern Physics}, 52(B), 274--284.\\
\noindent\small
\verb+https://doi.org/10.1016/j.shpsb.2015.08.010+
\normalsize

\bibspace
Mellor, D. H. (2005).
\textit{Probability: A Philosophical Introduction}.
Routledge.

\bibspace
Pais, A. (1986).
\textit{Inward Bound: Of Matter and Forces in the
Physical World}.
Oxford University Press.

\bibspace
Peierls, R. (1991).
In Defence of `Measurement'.
\textit{Physics World}, 4(1), 19--20.\\
\noindent\small
\verb+https://iopscience.iop.org/article/10.1088/2058-7058/4/1/19/meta+
\normalsize

\bibspace
Popper, K. R. (1959).
The Propensity Interpretation of Probability.
\textit{The British Journal for the Philosophy of
Science}, 10(37), 25--42.\\
\noindent\small
\verb+https://www.journals.uchicago.edu/doi/abs/10.1093/bjps/X.37.25?jour+\\
\verb+nalCode=bjps+
\normalsize

\bibspace
Popper, K. R. (1982).
\textit{Quantum Theory and the Schism in Physics}.
Hutchinson.

\bibspace
Rutherford. E. (1900).
A Radio-active Substance Emitted from
Thorium Compounds.
\textit{Philosophical Magazine}, 49(296), 1--14.\\
\noindent\small
\verb+https://www.tandfonline.com/doi/abs/10.1080/14786440009463821?journ+\\
\verb+alCode=tphm16+
\normalsize

\bibspace
Vaidman, L. (1998).
On Schizophrenic Experiences of the Neutron
or Why We Should Believe in the Many-Worlds
Interpretation of Quantum Theory.
\textit{International Studies in the Philosophy
of Science}, 12(3), 245--261.\\
\noindent\small
\verb+https://www.tandfonline.com/doi/abs/10.1080/02698599808573600+
\normalsize

\bibspace
Von Neumann, J. (1932).
\textit{Mathematical Foundations of Quantum Mechanics}.
(English translation by R. T. Beyer.)
Princeton University Press, 1955.

\bibspace
Wigner, E. P. (1961).
Remarks on the Mind-Body Question.
In \textit{The Scientist Speculates}
(edited by I. J. Good.).
William Heinemann, 284--302.
\end{document}